\documentclass[epj]{svjour}

\pdfoutput=1	

\usepackage{amsmath}	
\usepackage{graphicx}
\usepackage[outdir=../]{epstopdf}	
\usepackage{multirow}
\usepackage[pagebackref=false]{hyperref}	
\hypersetup{
  colorlinks = true,
  linkcolor=blue,   
  citecolor=blue,   
  urlcolor=blue,    
  pdfdisplaydoctitle=true
}
\usepackage[numbers, square, comma, sort&compress, merge]{natbib}	

\newcommand{\gosia}{\textsc{gosia}}	

\graphicspath{{../}}		

\begin{document}

\title{Low-energy Coulomb excitation of $^{62}$Fe and $^{62}$Mn following in-beam decay of $^{62}$Mn}
\titlerunning{Coulomb excitation of $^{62}$Fe and $^{62}$Mn}
\authorrunning{L.~P.~Gaffney et al.}

\author{
L.~P.~Gaffney\inst{1,}\thanks{\emph{Corresponding author:} Liam.Gaffney@uws.ac.uk}$^{,}$\thanks{\emph{Present address:} School of Engineering, University of the West of Scotland, Paisley PA1 2BE, United Kingdom} \and
J.~Van~de~Walle\inst{1,2} \and
B.~Bastin\inst{1,3} \and
V.~Bildstein\inst{4,}\thanks{\emph{Present address:} Department of Physics, University of Guelph, Guelph, ON, Canada N1G 2W1} \and
A.~Blazhev\inst{5} \and
N.~Bree\inst{1} \and
J.~Cederk\"{a}ll\inst{6} \and
I.~Darby\inst{7} \and
H.~De~Witte\inst{1} \and
D.~DiJulio\inst{6} \and
J.~Diriken\inst{1,8} \and
V.~N.~Fedosseev\inst{2} \and
Ch.~Fransen\inst{5} \and
R.~Gernh\"{a}user\inst{5} \and
A.~Gustafsson\inst{2} \and
H.~Hess\inst{5} \and
M.~Huyse\inst{1} \and
N.~Kesteloot\inst{1,8} \and
Th.~Kr\"{o}ll\inst{9} \and
R.~Lutter\inst{10} \and
B.~A.~Marsh\inst{2} \and
P.~Reiter\inst{5} \and
M.~Seidlitz\inst{5} \and
P.~Van~Duppen\inst{1} \and
D.~Voulot\inst{2} \and
N.~Warr\inst{5} \and
F.~Wenander\inst{2} \and
K.~Wimmer\inst{4,}\thanks{\emph{Present address:} Department of Physics, University of Tokyo, Hongo, Bunkyo-ku, Tokyo, 113-0033, Japan}\and
K.~Wrzosek-Lipska\inst{1,11}
}

\institute{
KU Leuven, Instituut voor Kern- en Stralingsfysica, 3001 Leuven, Belgium \and
CERN-ISOLDE, CERN, CH-1211 Geneva 23, Switzerland \and
GANIL CEA/DSM-CNRS/IN2P3, Boulevard H. Becquerel, F-14076 Caen, France \and
Physics Department E12, Technische Universit\"{a}t M\"{u}nchen, D-85748 Garching, Germany \and
Institut f\"{u}r Kernphysik, Universit\"{a}t zu K\"{o}ln, 50937 K\"{o}ln, Germany \and
Physics Department, University of Lund, Box 118, SE-221 00 Lund, Sweden \and
University of Jyvaskyla, Department of Physics, FI-40014 University of Jyvaskyla, Finland \and
Belgian Nuclear Research Centre SCK$\bullet$CEN, Boeretang 200, B-2400 Mol, Belgium \and
Institut f\"{u}r Kernphysik, Technische Universit\"{a}t Darmstadt, D-64289 Darmstadt, Germany \and
Ludwig-Maximilians-Universit\"{a}t-M\"{u}nchen, Schellingstra{\ss}e 4, 80799 M\"{u}nchen, Germany \and
Heavy Ion Laboratory, University of Warsaw, PL-00-681 Warsaw, Poland
}

\date{Published: 28$^{\mathrm{th}}$ October 2015}

\abstract{
Sub-barrier Coulomb-excitation was performed on a mixed beam of $^{62}$Mn and $^{62}$Fe, following in-trap $\beta^{-}$ decay of $^{62}$Mn at REX-ISOLDE, CERN.
The trapping and charge breeding times were varied in order to alter the composition of the beam, which was measured by means of an ionisation chamber at the zero-angle position of the Miniball array.
A new transition was observed at 418~keV, which has been tentatively associated to a $(2^{+},3^{+})\rightarrow1^{+}_{g.s.}$ transition.
This fixes the relative positions of the $\beta$-decaying $4^{+}$ and $1^{+}$ states in $^{62}$Mn for the first time.
Population of the $2^{+}_{1}$ state was observed in $^{62}$Fe and the cross-section determined by normalisation to the $^{109}$Ag target excitation, confirming the $B(E2)$ value measured in recoil-distance lifetime experiments.
} 

%
\maketitle

\section{Introduction}

Shell-closures in nuclei are known to evolve when moving away from the line of $\beta$ stability, with vanishing or newly appearing energy gaps~\cite{Thibault1975,Poves1987,Fukunishi1992,Otsuka2001,Warner2004,Bastin2007,Sorlin2008}.
At $N=20$, this changing shell-structure leads to an ``island of inversion''~\cite{Warburton1990,Sorlin2008}, whereby the usual spherical mean-field gap between the $sd$ and $fp$ shells is eroded by strong quadrupole correlations.
As a consequence, deformed intruder states become the major component of the ground state around $^{32}$Mg~\cite{Pritychenko1999}.
The spherical structures then appear as excited $0^{+}$ states in the even-even members of this so-called island~\cite{Wimmer2010}, leading to what can also be described as a type of shape coexistence~\cite{Heyde2011}.
This unexpected increase of quadrupole collectivity at shell closures is not only restricted to the ${N=20}$ region, however.
Indicated by the high-lying $2^{+}_{1}$ state and low ${B(E2; 2^{+}_{1} \to 0^{+}_{1})}$ value in $^{68}$Ni~\cite{Sorlin2002,Bree2008}, the weak ${N=40}$ sub-shell gap vanishes with the removal or addition of only a couple of protons~\cite{Hannawald1999,Gade2010}.
Excited $0^{+}$ states have been identified in $^{68}$Ni~\cite{Elseviers2014,Recchia2013,Suchyta2014} and are suggested to be deformed intruder configurations in recent Monte Carlo Shell Model (MCSM) calculations~\cite{Tsunoda2014}.
The coexistence of these shapes~\cite{Heyde2011,Tsunoda2014,Suchyta2014} leads to the proposition of a new island of inversion around $^{64}$Cr, described in shell-model calculations using the new LNPS effective interaction~\cite{Lenzi2010}.
In the Cr isotopic chain with ${32<N<40}$, a gradual onset of quadrupole collectivity is observed~\cite{Burger2005,Seidlitz2011a,Baugher2012,Crawford2013,Braunroth2015}.
Recent $B(E2)$ measurements were performed in the neutron-rich Fe isotopes using the recoil-distance lifetime technique~\cite{Ljungvall2010,Rother2011} and intermediate energy Coulomb excitation~\cite{Crawford2013}, showing a sudden increase in deformation at $N=38$.
In this experiment, low-energy Coulomb excitation of $^{62}$Fe can provide a confirmation of this observation.

In the odd-mass and odd-$Z$ nuclei about $^{64}$Cr, the role of the $\nu g_{9/2}$ orbital can be mapped more directly.
Shell-model calculations~\cite{Srivastava2010} have shown the importance of this intruder orbital as neutron excitations across $N=40$ play a key role in the most neutron-rich Mn isotopes.
Recent experiments have identified negative-parity structures associated with occupancy of the $\nu g_{9/2}$ orbital~\cite{Chiara2010} in a number of these nuclei.
However, the behaviour of these states with respect to the ground-state configuration is limited in $^{62}$Mn due to the unknown positioning of the isomer and ground-state levels.
It has recently been confirmed that the two $\beta$-decaying states have $I^{\pi}=1^{+}$ and $4^{+}$~\cite{Heylen2015a}, but their relative positions remain unknown.
With the Coulomb-excitation technique, it has been possible in this paper to identify a new state built upon the longer-lived $4^{+}$ state in $^{62}$Mn.
Here we report on its identification and propose the ordering and relative energies of the two $\beta$-decaying states; in turn we also fix the energy of the negative parity states identified in Ref.~\cite{Chiara2010}.

\section{Experimental Details}

Extraction of refractory and refractory-like elements in atomic form from ISOL (Isotope Separator On-Line) targets is very slow and inefficient.
The atoms first need to diffuse through the target matrix to the surface and then desorb before reaching the ion source.
Many elements will undergo this process quickly and may collide with the walls of the heated transfer line ($\approx2000^{\circ}$C) many times en route to the ion source, adsorbing and desorbing quickly each time.
The biggest bottle neck in the extraction of refractory-like elements, such as Fe, comes from slow desorption~\cite{Kirchner1992}.
This may be seen as a benefit where isobaric purity is required, but certainly not when beams of short-lived isotopes of such troublesome elements are desired.
Fast and chemically-independent extraction is planned at future facilities with new gas-catcher and laser-ionisation coupled systems~\cite{Kudryavtsev2013}.
Gas-catchers are currently employed at the CARIBU experiment at ANL~\cite{caribu}, producing beams of Cf fission products that have been reaccelerated for Coulomb-excitation studies.
With the thick solid target at ISOLDE however, alternative techniques than direct extraction have to be explored, such as forming volatile gaseous compounds~\cite{Koster2007}.
Recently, the technique of in-trap decay was pioneered at REX-ISOLDE using $^{61}$Mn as a test case to produce a beam of the refractory-like $^{61}$Fe~\cite{VandeWalle2009}.
Following these successful tests, a further experiment on $^{62}$Mn/Fe was performed and is described here.

The manganese beams are produced at ISOLDE by proton-induced fission of uranium, driven by a 1.4~GeV proton beam from the CERN's PS Booster, impinging on a thick UC$_{x}$ target.
Element-selective ionisation takes place with the Resonance Laser Ionisation Laser Ion Source (RILIS)~\cite{Fedoseyev1997,Hannawald1999,Fedosseev2012} and the Mn$^{1+}$ ions are extracted from the ion source with a 30~kV potential and mass separated with the General Purpose Separator (GPS).
After this the beam is bunched in REX-TRAP~\cite{Wolf2003,Wenander2010} and released into the charge breeder, REX-EBIS~\cite{Wolf2003,Wenander2010}, finally attaining a charge state of $21^{+}$.
The latter stage is required in order to achieve a mass to charge ratio of $A/q<4.5$ for injection to the REX linear accelerator~\cite{Voulot2008}.
The distribution of charge states is sensitive to the charge-breeding time.
Both the trapping and charge-breeding times were varied during the experiment in order to investigate the production of $^{62}$Fe via the ``in-trap'' method (see Section~\ref{sec:intrapdecay}).
Ultimately, the beam was delivered to the Miniball array~\cite{Warr2013} at a final energy of 2.86~MeV/$u$.
Here it was either incident on a 4~mg/cm$^{2}$-thick $^{109}$Ag target to induce Coulomb excitation, or into the ionisation chamber, filled with CF$_{4}$ gas and backed by a thick silicon detector, to monitor the beam composition~\cite{Warr2013}.

\subsection{In-trap decay} \label{sec:intrapdecay}

\begin{figure}[tb]
\centering
\includegraphics[width=\columnwidth]{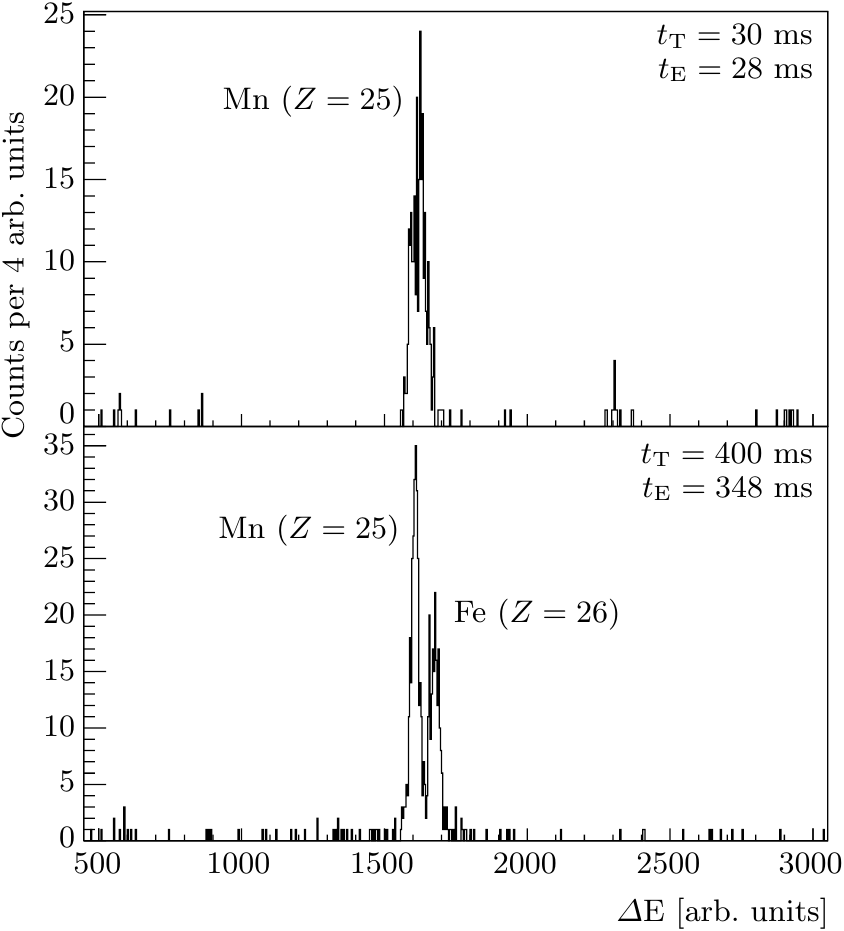}
\caption{${\Delta}E$ signal as measured in arbitrary units from the ionisation chamber, proportional to energy loss of the beam, ${\Delta}E$, for two different trapping and charge-breeding settings.
Top panel: 30~ms trapping time, $t_{\mathrm{T}}$, and 28~ms charge-breeding time, $t_{\mathrm{E}}$.
Bottom panel: $t_{\mathrm{T}}=400$~ms and $t_{\mathrm{E}}=348$~ms.}
\label{fig:ionchamber}
\end{figure}

\begin{figure}[tb]
\centering
\includegraphics[width=\columnwidth]{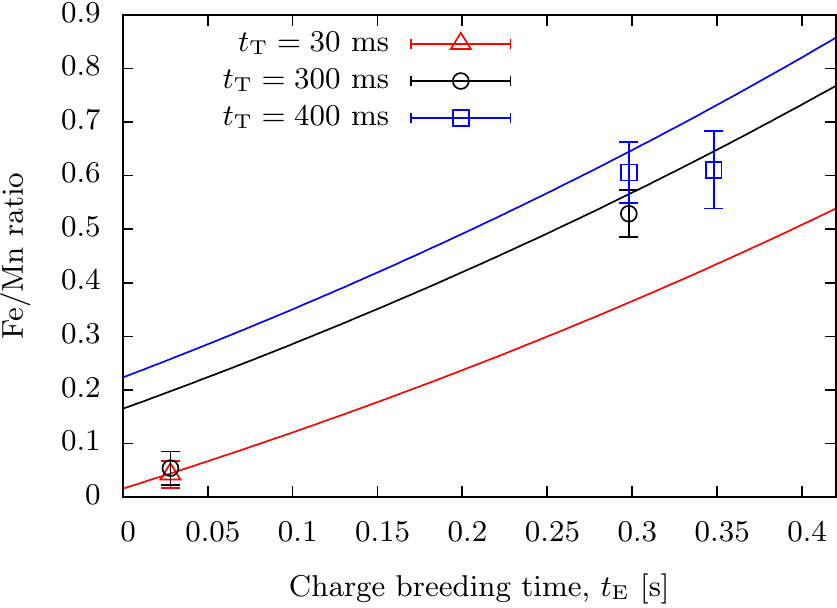}
\caption{Ratio of $^{62}$Fe to $^{62}$Mn in the beam, measured experimentally in the ionisation chamber, as a function of charge breeding time, $t_{\mathrm{E}}$. Theoretical curves for different trapping times, $t_{\mathrm{T}}$, are also shown in red, black and blue corresponding to 30~ms, 300~ms and 400~ms, respectively. It is assumed that the halflife of the $^{62}$Mn component is 671~ms, corresponding to the high-spin $\beta$-decaying state~\cite{Hannawald1999}.}
\label{fig:beamratio}
\end{figure}

During the trapping and charge-breeding stages, the $^{62}$Mn ions are held for varying time periods from 28~ms to 740~ms.
At a number of these timing settings, the beam composition was measured in the $\Delta{}E_{\mathrm{gas}}$-$E_{\mathrm{Si}}$ telescope of the ionisation chamber. An example of the $\Delta{}E_{\mathrm{gas}}$ spectra are shown in Fig.~\ref{fig:ionchamber}.
The energy loss in the gas is proportional to the $Z$ of the projectile and is sensitive to the gas conditions such as pressure.
Good optimisation of these conditions allowed for a clear separation of $Z=25,26$ and therefore the ratio of manganese to iron in the beam.
Under idealised conditions, i.e. no losses during trapping and charge breeding and a constant rate of injection to REX-TRAP, this ratio of manganese and iron can be calculated using simple mother-daughter decay laws.
Assuming the injection of the high-spin state of $^{62}$Mn(T$_{1/2}=671(5)$~ms~\cite{Hannawald1999}) into the trap (see Section~\ref{sec:418keV}) at a constant rate, its decay and the grow-in and decay of $^{62}$Fe(T$_{1/2}=68(2)$~s~\cite{Franz1975}) can be integrated over the total trapping time, $t_{\mathrm{T}}$.
\begin{equation}
\frac{N^{\mathrm{T}}_{\mathrm{Mn}}}{N_{0}} = \frac{1}{t_{\mathrm{T}}}
\int^{t_{\mathrm{T}}}_{0} e^{-\lambda_{\mathrm{Mn}} t } ~dt	,
\end{equation}\begin{equation}
\frac{N^{\mathrm{T}}_{\mathrm{Fe}}}{N_{0}} = \frac{1}{t_{\mathrm{T}}}
\int^{t_{\mathrm{T}}}_{0} \frac{ \lambda_{\mathrm{Mn}} }{ \lambda_{\mathrm{Fe}} - \lambda_{\mathrm{Mn}} }
\left( e^{-\lambda_{\mathrm{Mn}} t }  -  e^{-\lambda_{\mathrm{Fe}} t } \right) ~dt	,
\end{equation}
where $N_{0}$ and $N^{\mathrm{T}}_{\mathrm{Mn}}$ are the numbers of Mn$^{+}$ ions injected into and extracted from the trap, respectively, and $N^{\mathrm{T}}_{\mathrm{Fe}}$ is the number of Fe$^{+}$ ions extracted from the trap. $N^{\mathrm{T}}_{\mathrm{Mn}}$ and $N^{\mathrm{T}}_{\mathrm{Fe}}$ are then treated as the number of the respective ion species that are injected into the EBIS charge-state breeder, where further decay occurs while the beam is confined for the charge breeding time, $t_{\mathrm{E}}$.
\begin{equation}
\frac{N^{\mathrm{E}}_{\mathrm{Mn}}}{N_{0}} =
\frac{N^{\mathrm{T}}_{\mathrm{Mn}}}{N_{0}}
e^{ - \lambda_{\mathrm{Mn}} t_{\mathrm{E}} }	,
\end{equation}
\begin{equation}
\begin{split}
\frac{N^{\mathrm{E}}_{\mathrm{Fe}}}{N_{0}} = & 
\frac{N^{\mathrm{T}}_{\mathrm{Fe}}}{N_{0}} e^{ - \lambda_{\mathrm{Fe}} t_{\mathrm{E}} } \\
& + \frac{N^{\mathrm{T}}_{\mathrm{Mn}}}{N_{0}} \frac{ \lambda_{\mathrm{Mn}} }{ \lambda_{\mathrm{Fe}} - \lambda_{\mathrm{Mn}} }
\left( e^{-\lambda_{\mathrm{Mn}} t_{\mathrm{E}} }  -  e^{-\lambda_{\mathrm{Fe}} t_{\mathrm{E}} } \right)	.
\end{split}
\end{equation}
Figure~\ref{fig:beamratio} shows the calculated ratio of $^{62}$Fe and $^{62}$Mn for different trapping times, compared to the experimentally determined values.
One can see that the behaviour is generally well understood and an appreciable amount of iron can be accumulated in the beam.
For the longest trapping times, the iron-to-manganese content is generally less than expected.
This may be due in part to space-charge effects, which limit the number of ions that can be fed into the trap in each bunch.
Furthermore, the recoil energy of the Fe ions following the $\beta$-decay process may cause losses in the trap.
This is highlighted by the data point at $t_{\mathrm{T}}=300$~ms and $t_{\mathrm{E}}=28$~ms that lies close to the value taken with a much shorter trapping time of $t_{\mathrm{T}}=30$~ms. 
While it appears as though longer charge breeding times are preferable to increase the Fe/Mn ratio, the efficiency of achieving the correct charge state for injection to the REX accelerator reduces and an overall loss of beam intensity begins to take over.
Additionally, the charge distributions of Mn and Fe ions in REX-EBIS will be different meaning that these effects cannot be separated.

\subsection{Coulomb excitation}

\begin{figure}[tb]
\centering
\includegraphics[width=0.96\columnwidth]{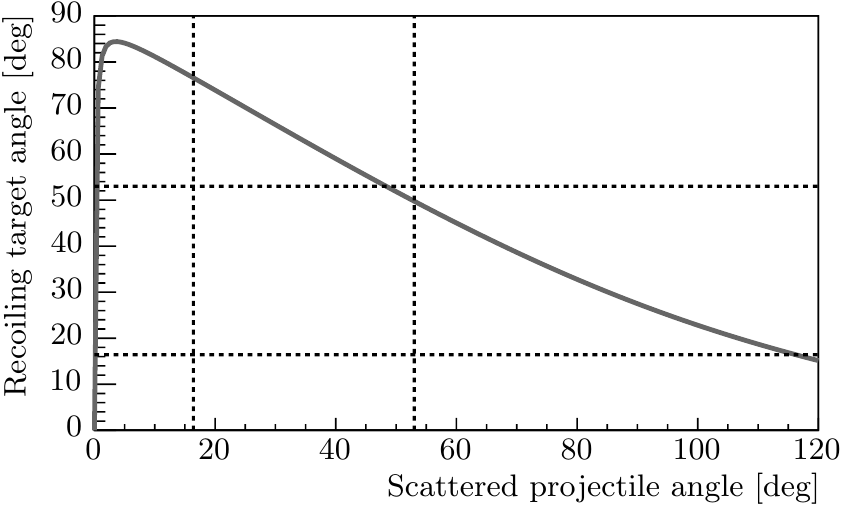}
\caption{Kinematic relationship of laboratory scattering angles for a mass $A=62$ projectile impinging on a mass $A=109$ target at an energy of 2.86~MeV/$u$. It is assumed that there is an excitation of 0.877~MeV. The horizontal and vertical dashed lines represent the limits of detection in the CD detector. Only for a small range of scattering angles are both the projectile and recoil expected to be in the range of detection simultaneously.}
\label{fig:kinematics}
\end{figure}

With the $^{109}$Ag target in place, ``safe'' Coulomb excitation~\cite{Cline1986} takes place and the scattered projectiles and recoiling target nuclei are detected in a compact-disc-shaped Double-Sided Silicon Strip Detector (DSSSD), the so-called CD detector, placed 32.5~mm downstream from the target position~\cite{Warr2013}.
This corresponds to an angular coverage in the lab frame of $15.6^{\circ}$--$51.6^{\circ}$ and in the centre-of-mass (CoM) frame of $24.3^{\circ}$--$148.9^{\circ}$.
For the evaluation of the cross-section the inner-most strip is not used since it was broken on one of the four quadrants.
The de-excitation $\gamma$ rays are detected in the HPGe crystals of Miniball, which each have six-fold segmentation, arranged into eight triple clusters. 
Prompt and random coincidences between particle and $\gamma$-ray events are built in the manner described in Ref.~\cite{Gaffney2015-Rn}.
Furthermore, for angles greater than $\approx48^{\circ}$ in the lab frame, both scattering partners for a given CoM scattering angle will be inside of the CD range (see Figure~\ref{fig:kinematics}). Therefore, two-particle coincidences in this range are also considered in the analysis.

\begin{figure}[tb]
\centering
\def\svgwidth{0.98\columnwidth}
\includegraphics[width=\columnwidth]{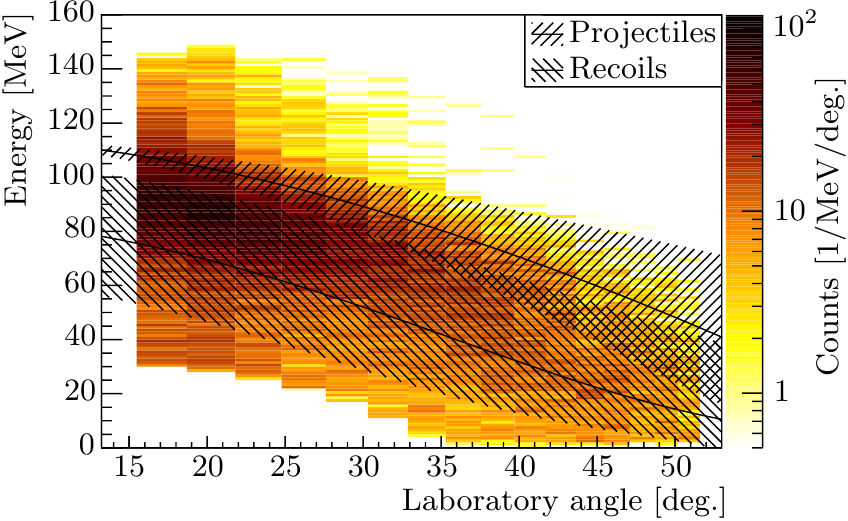}
\caption{
(Color online) Detected particle events in the CD detector plotted as a function of angle and energy in the laboratory frame of reference.
An angle-dependent energy threshold has been applied in software to remove low-energy events caused by noise on the detector.
The two shaded regions show the calculated behaviour of the projectile and recoils for elastic scattering including energy loss in the target.
The range is determined by the assumption of the reaction at the entrance or exit of the target, while the solid line assumes a reaction in the centre.
However, the detector resolution is not considered here, nor is the finite width of the CD-detector strips.
}
\label{fig:particles}
\end{figure}

A relatively thick target (4.0~mg/cm$^{2}$) was used in order to increase the yield, which resulted in poor energy resolution of the scattered particles.
In turn, it was not possible to cleanly resolve the scattered projectiles and target recoils in terms of energy and laboratory angle, as can be seen in Figure~\ref{fig:particles}.
A heavier-mass target species that still provides the relative normalisation via observable excitation would have allowed a greater separation of the two kinematic solutions in energy.
Nevertheless, under the assumption that either a projectile-like or recoil-like particle is detected, it is possible to solve the two-body kinematic problem for the other scattering partner.
It appears as though the detected energy is systematically lower than that expected from the calculations (see Fig.~\ref{fig:particles}).
The calibration of the detector used a triple-$\alpha$ source of $^{239}$Pu-$^{241}$Am-$^{243}$Cm with energies around 5~MeV and the extrapolation to higher energies may well be non-linear.
At the velocities used in this experiment, the Doppler shift of $\gamma$-rays emitted in flight is dominated by the angular component and the correction is rather insensitive to changes in the particle energy.

\begin{figure}[tb]
\centering
\includegraphics[width=\columnwidth]{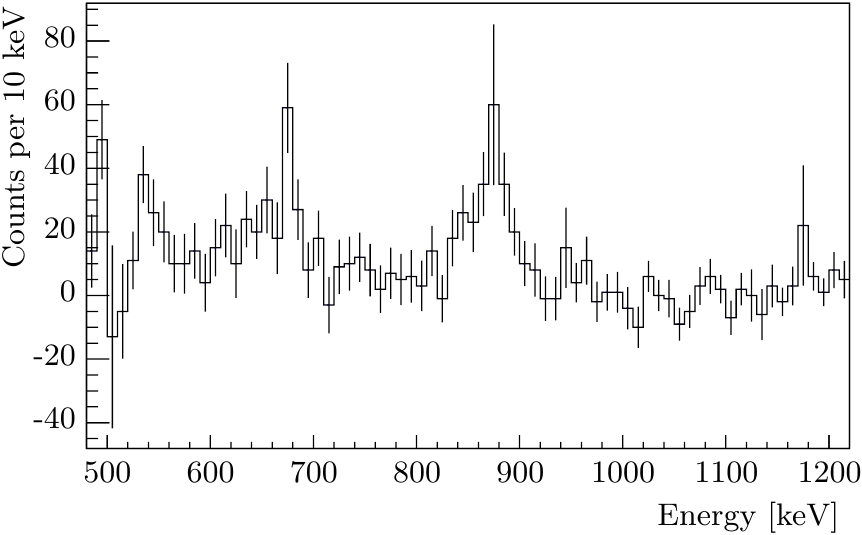}
\caption{
Background-subtracted $\gamma$-ray spectrum showing the 877~keV ${2^{+}_{1} \to 0^{+}_{1}}$ transition of $^{62}$Fe following Coulomb excitation on a 4.0~mg/cm$^{2}$-thick $^{109}$Ag target.
Data at all trapping and charge-breeding settings are summed in this spectrum, although 75\% data was taken with the longest times.
Note here that no Doppler correction is applied because of the ambiguity of projectile- and recoil-like events.
This leads to the wide peak with an irregular (non-Gaussian) peak shape.
The structure at $\approx670$~keV is likely to be the Compton edge of the 877~keV peak.
}
\label{fig:gammas}
\end{figure}

A background subtraction of the $\gamma$-ray spectra is performed using the randomly-coincident events, normalised to the intensity of $\gamma$ rays emitted following the $\beta$-decay of $^{62}$Fe and $^{62}$Co.
The background subtraction is key since the $\beta$ decay of the $^{62}$Mn mother (scattered into the CD detector and the walls of the chamber) will populate the $2^{+}_{1}$ state in $^{62}$Fe, giving rise to false coincidences with non-Coulomb-excitation-related ${2^{+}_{1} \to 0^{+}_{1}}$ $\gamma$-rays.
As can be observed in the background-subtracted $\gamma$-ray spectrum of Fig.~\ref{fig:gammas}, the uncertainty on the number of counts in the region around 877~keV (also 511~keV associated with the annihilation peak) is large because of this subtraction.
An additional 8.1\% uncertainty is calculated from the uncertainty in the determination of the background normalisation.
All of this propagates to the total uncertainty in the integrated intensity of the ${2^{+}_{1} \to 0^{+}_{1}}$ (shown in Table~\ref{tab:yields}), hindering the precision of the cross-section measurement.

\begin{table}[tb]
\caption{Experimental intensities of $\gamma$-ray transitions following Coulomb excitation of the mixed $^{62}$Mn/Fe beam on a $^{109}$Ag target.
A correction for the relative efficiency of the Miniball array, normalised to 877.3~keV, along with its uncertainty is included.
The assignment of the 418~keV transition is discussed in Section~\ref{sec:418keV}.}
\label{tab:yields}
\centering
\begin{tabular}{cccc}
\hline\noalign{\smallskip}
 & Energy [keV] & Transition & $I_{\gamma}$  \\
\noalign{\smallskip}\hline\noalign{\smallskip}
$^{62}$Fe & 877.3 & ${2^{+}_{1} \to 0^{+}_{1}}$ & 210(40) \\
$^{109}$Ag & 311.4 & ${3/2^{-}_{1} \to 1/2^{-}_{1}}$ & 1520(40) \\
$^{109}$Ag & 415.2 & ${5/2^{-}_{1} \to 1/2^{-}_{1}}$ & 2300(60) \\
$^{62}$Mn & 418(2) & ${(2^{+},3^{+}) \to 1^{+}_{1}}$ & 970(50) \\
\noalign{\smallskip}\hline
\end{tabular}
\end{table}

\begin{figure}[tb]
\centering
\includegraphics[width=0.98\columnwidth]{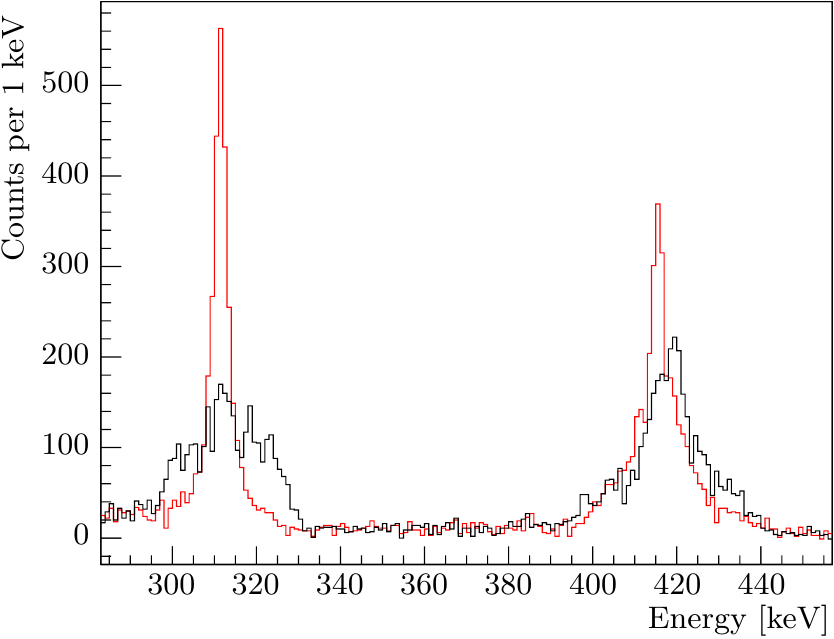}
\caption{(Color online) Background-subtracted $\gamma$-ray spectra showing the region of the target de-excitation.
Data at all trapping and charge-breeding settings are summed in these spectra, although ~75\% data was taken with the longest times.
The spectra are Doppler corrected assuming detection of a projectile-like particle and $\gamma$-ray emission from either a projectile-like (black) or recoil-like (red) particle.
The line shape of the 311~keV peak is symmetric and as expected for a single $\gamma$-ray transition from the recoil.
The 415~keV peak clearly shows contamination, and when Doppler corrected for the projectile, a component at 418~keV emerges.}
\label{fig:targetlines}
\end{figure}

In order to extract the cross-section, we normalise to the excitation of the $^{109}$Ag target. At this point, two issues are encountered.
First of all, both components of the beam, $^{62}$Fe and $^{62}$Mn, will give rise to excitation of the target.
With knowledge of the total ratio of manganese to iron throughout the experiment, $R_{\mathrm{Mn}}=\frac{N_{\mathrm{Mn}}}{N_{\mathrm{Fe}}}=1.7(2)$, the fraction of the observed de-excitation intensity in the target associated with $^{62}$Fe can be deduced~\cite{Zielinska2015}:
\begin{equation} \label{eq:beamimpurities}
F_{\mathrm{Fe}} = \frac{1}{ 1 + \left( R_{\mathrm{Mn}} \frac{\sigma_{t}\left({}^{62}\mathrm{Mn}\right)}{\sigma_{t}\left({}^{62}\mathrm{Fe}\right)} \right) } = 0.36(3)	,
\end{equation}
where $\frac{\sigma_{t}({}^{62}\mathrm{Fe})}{\sigma_{t}({}^{62}\mathrm{Mn})}$ and is the ratio of cross sections of exciting the state of interest in the target, by a $^{62}$Fe or $^{62}$Mn beam.
Calculated using the Coulomb-excitation code, \gosia{}~\cite{Czosnyka1983,GosiaManual}, this ratio is 0.9915 for the $3/2^{-}_{1}$ state at 311~keV in $^{109}$Ag.
Secondly, there is an additional component in the Doppler-broadened $\gamma$-ray peak of the ${5/2^{-}_{1} \to 1/2^{-}_{1}}$ transition in $^{109}$Ag around 415~keV.
Using the differences between peak shape in the Doppler-corrected spectra of Figure~\ref{fig:targetlines} a second component at 418~keV emerges, associated with the projectile.
Its intensity can be extracted by subtracting the true ${5/2^{-}_{1} \to 1/2^{-}_{1}}$-transition intensity extracted from the ratio, $\frac{I_{\gamma}(5/2^{-}_{1} \to 1/2^{-}_{1})}{I_{\gamma}(3/2^{-}_{1} \to 1/2^{-}_{1})}=1.13$, calculated in \gosia{} using previously measured matrix elements, and the ``clean'' intensity of the 311~keV, ${3/2^{-}_{1} \to 1/2^{-}_{1}}$ transition.
The deduced intensity of the 418~keV transition is almost five times greater than the ${2^{+}_{1} \to 0^{+}_{1}}$ transition in $^{62}$Fe (see Table~\ref{tab:yields}), meaning that this must originate from $^{62}$Mn. The assignment of this transition will be discussed in Section~\ref{sec:418keV}.
A search of particle-$\gamma$-$\gamma$ events showed no evidence of other transitions in coincidence with the new 418~keV transition.

\section{Results and discussion}

\subsection{Excitation of the $2^{+}_{1}$ state in $^{62}$Fe}

Contamination of the ${5/2^{-}_{1} \to 1/2^{-}_{1}}$ transition in the $^{109}$Ag target meant that normalisation could only be performed with respect to the ${3/2^{-}_{1} \to 1/2^{-}_{1}}$ transition.
A complete set of $E2$ and $M1$ matrix elements coupling the low-lying levels of $^{109}$Ag ($1/2^{-}_{1}$, $3/2^{-}_{1}$, $5/2^{-}_{1}$) are defined in the special version of the \gosia{} code, \gosia{}2~\cite{Czosnyka1983,GosiaManual}.
Additionally, known matrix elements to the higher-lying states ($3/2^{-}_{2}$, $5/2^{-}_{2}$) were included and used as buffer states in the calculation.
The \gosia{}2 version of the code allows for a simultaneous least-squares fit of matrix elements in the target and projectile systems, as well as a common set of normalisation constants, to the observed $\gamma$-ray intensities~\cite{Zielinska2015}.
The target system is over-determined by a complete set of spectroscopic data~\cite{Robinson1970,Throop1972,Miller1974,Loiselet1989,Blachot2006,Zielinska2009a} plus the observed ${3/2^{-}_{1} \to 1/2^{-}_{1}}$ intensity, which therefore provides the constraint on the normalisation constant.
However, the projectile system is under-determined.

\begin{figure}[tb]
\centering
\includegraphics[width=0.99\columnwidth]{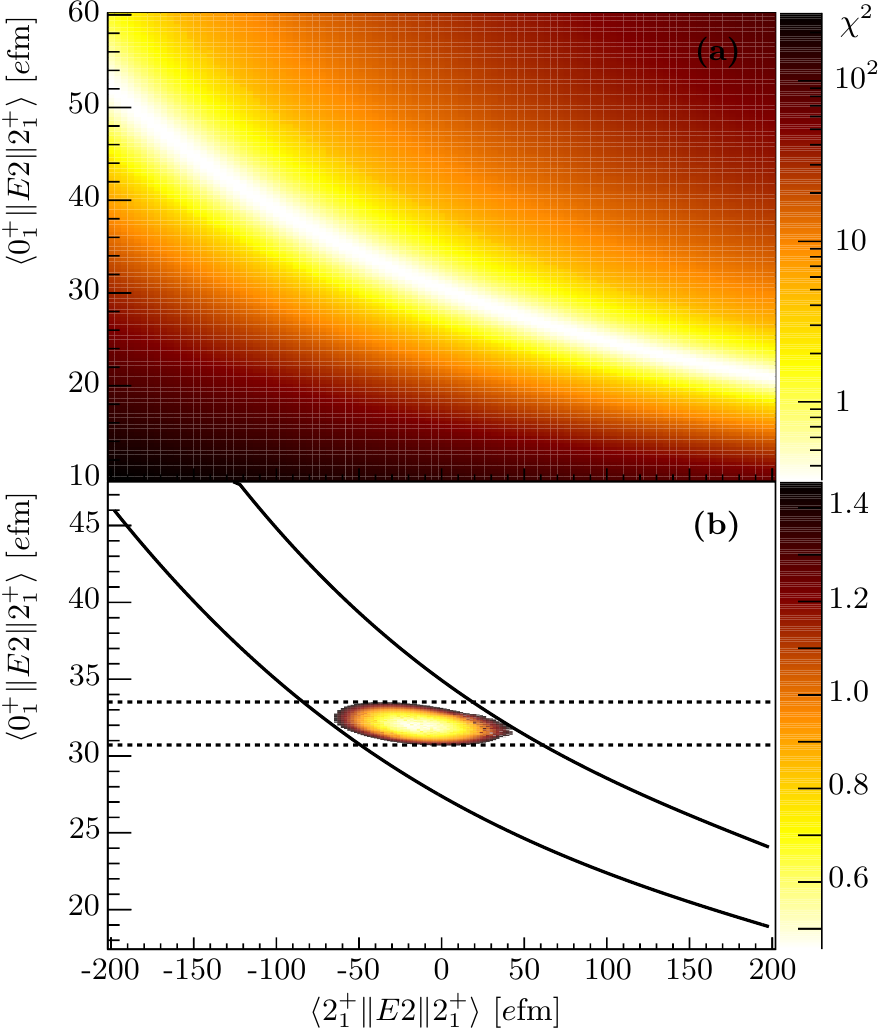}
\vspace{2mm}
\caption{(Color online)
(a) A full two-dimensional total-$\chi^{2}$ surface with respect to $\langle 2^{+}_{1} \|E2\| 0^{+}_{1} \rangle$ and $\langle 2^{+}_{1} \|E2\| 2^{+}_{1} \rangle$ for the $^{62}$Fe projectile.
(b) The resulting surface when combined with the lifetime measurements of Refs.~\cite{Ljungvall2010,Rother2011} and a cut applied with the condition that $\chi^{2}<\chi^{2}_{\mathrm{min}}+1$, representing $1\sigma$. The individual $1\sigma$ contours for the Coulomb-excitation and lifetime data are shown by the solid and dashed lines, respectively.}
\label{fig:chisqsurface}
\end{figure}

In $^{62}$Fe, the excitation cross-section of the $2^{+}_{1}$ is governed in this experiment by only two matrix elements, namely ${\langle 0^{+}_{1} \| E2 \| 2^{+}_{1} \rangle}$ and ${\langle 2^{+}_{1} \| E2 \| 2^{+}_{1} \rangle}$.
All other matrix elements can be shown to influence at the level of less than 1\%.
However, only one transition intensity is experimentally determined, which means that there is no unique solution to the system.
Usually, segmentation of the data in different ranges of CoM scattering angle provides additional data~\cite{Zielinska2015,Gaffney2015-Rn}, but because of the ambiguity in the particle kinematics (see Figure~\ref{fig:particles}) and low statistics, it is not possible in this case.
Instead, a total-$\chi^{2}$ surface is created in two dimensions, representing each matrix element, including the sum of the contributions from the target and projectile systems and is plotted in Figure~\ref{fig:chisqsurface}(a).
Recent lifetime measurements utilising the recoil-distance Doppler-shift method, by Ljungvall~et~al.~\cite{Ljungvall2010} and Rother~et~al.~\cite{Rother2011}, provide a further constraint on the total-$\chi^{2}$ surface, represented by the dashed lines in Figure~\ref{fig:chisqsurface}(b).
In order to treat the uncertainties correctly, both of these lifetime data were included in the \gosia{}2 fit and treated with equal weight to the $\gamma$-ray intensities and other data.
The resulting 1$\sigma$ surface is now constrained in both dimensions resulting in an ellipse that represents the correlations between the two parameters.
By projecting the point with the minimum $\chi^{2}$ and the limits of this ellipse onto the relevant axis, we are able to extract values and uncertainties of the matrix elements, shown in Table~\ref{tab:results}.
If one were to assume that the spectroscopic quadrupole moment, $Q_{s}(2^{+}_{1})$ is zero, or otherwise that the second-order excitation process can be neglected, a projection can be made to ${\langle 0^{+}_{1} \| E2 \| 2^{+}_{1} \rangle}$ at ${\langle 2^{+}_{1} \| E2 \| 2^{+}_{1} \rangle = 0}$.
The values obtained from this procedure without including the additional lifetime data (i.e projecting Figure~\ref{fig:chisqsurface}(a) at $x=0$) are also presented in Table~\ref{tab:results}, although the uncertainties should be considered to be underestimated by neglecting correlations.
Consistency between the lifetimes and the current Coulomb-excitation data is however shown.

\begin{table}
\caption{Results in $^{62}$Fe projected from the ellipse obtained using the full data set, including the lifetimes of Refs.~\cite{Ljungvall2010,Rother2011}, and projecting the $\chi^{2}$ surface at $Q_{s}(2^{+}_{1})=0$ using only the Coulomb-excitation data. Note that ${\langle 0^{+}_{1} \| E2 \| 2^{+}_{1} \rangle}$ and the corresponding $B(E2)$ value are constrained entirely by the lifetime data.}
\label{tab:results}       
\centering
\begin{tabular}{rcc}
\hline\noalign{\smallskip}
 & with $\tau(2^{+}_{1})$ & with $Q_{s}(2^{+}_{1})=0$  \\[1pt]
\hline\noalign{\smallskip}
${\langle 0^{+}_{1} \| E2 \| 2^{+}_{1} \rangle}=$ & $32.0^{+1.5}_{-1.3}$~$e$fm$^{2}$ & $31^{+4}_{-3}$~$e$fm$^{2}$ \\[2pt]
${\langle 2^{+}_{1} \| E2 \| 2^{+}_{1} \rangle}=$ & $-10^{+60}_{-50}$~$e$fm$^{2}$ & -- \\[2pt]
$B(E2; 2^{+}_{1} \to 0^{+}_{1})=$ & $14.0^{+1.3}_{-1.1}$~W.u. & $13^{+4}_{-3}$~W.u. \\[2pt]
 $Q_{s}(2^{+}_{1})=$ & $-8^{+40}_{-40}$~$e$fm$^{2}$ & -- \\
\noalign{\smallskip}\hline
\end{tabular}
\end{table}

Sensitivity to the sign of $Q_{s}$ is desirable in order to determine the nature and not only the magnitude of deformation.
While state-of-the-art shell-model calculations using the LNPS interaction predict a prolate deformation with $Q_{s}(2^{+}_{1})\approx-30$~$e$fm$^{2}$~\cite{Lenzi2010}, a relatively simple analysis of Nilsson orbitals with realistic Woods-Saxon potentials predicts an oblate deformation in $^{62}$Fe~\cite{Hamamoto2014}.
Unfortunately, the current level of uncertainty does not allow one to distinguish between the two possibilities.
In principle however, the sensitivity to $Q_{s}(2^{+}_{1})$ can be increased by the variation of scattering parameters such as beam energy, target species or scattering angle.
This experiment demonstrates the method and feasibility of any potential new measurements.
Higher-beam energies provided by HIE-ISOLDE will further increase the cross-section, increasing the statistical precision.
The potential for the emergence of shape coexistence in this region~\cite{Heyde2011,Carpenter2013} increases the interest in more precision measurements.

\subsection{Assignment of the 418~keV transition in $^{62}$Mn} \label{sec:418keV}

The identification, in this experiment, of a 418~keV $\gamma$-ray transition cannot be reconciled with previous experiments employing $\beta$-decay~\cite{Gaudefroy2005}, multi-nucleon transfer reactions~\cite{Valiente-Dobon2008}, or deep-inelastic reactions~\cite{Chiara2010}.
No such transition is observed in these data, although the proposed ${I=(6)}$ state, identified in both in-beam experiments~\cite{Valiente-Dobon2008,Chiara2010}, lies 418~keV above the $\beta$-decaying $4^{+}$ state.
However, the non-observation of the $(6)\to(5)$ 196~keV transition in our experiment rules out such a placement in the level-scheme.
To understand the origin of this transition, it must be known if the ground or the isomeric state is Coulomb-excited.
A number of experiments have been performed at ISOLDE recently that allow us to shed light on this question. 
Firstly, the spins of both $\beta$-decaying states were confirmed in a laser-spectroscopy experiment at the COLLAPS setup~\cite{Heylen2015,Heylen2015a}. In this experiment, the longer-lived $4^{+}$ state (T$_{1/2}=671(5)$~ms~\cite{Hannawald1999}) was observed to have been extracted with a much higher intensity than the $1^{+}$ state (T$_{1/2}=92(13)$~ms~\cite{Sorlin1999,Gaudefroy2005}).
Secondly, a $\beta$-decay experiment of $^{62}$Mn~\cite{Pauwels2014,NNDC} identified an excited $0^{+}$ state that is populated only in the $\beta$ decay of the shorter-lived $1^{+}$ state, giving rise to the 815~keV $\gamma$ ray observed in an earlier experiment~\cite{Gaudefroy2005}.

Coupling the observed extraction of the shorter-lived $1^{+}$ state in the COLLAPS experiment with the long trapping plus charge-breeding times (in excess of 700~ms for the most significant fraction of the Coulomb-excitation data), it is expected that only the long-lived state in $^{62}$Mn persists at the target position in this experiment.
A $\gamma$-$\gamma$ matrix was produced from the sum of all data collected during the ``beam-on'' and ``beam-off'' windows throughout the run.
The $0^{+}_{2}\to2^{+}_{1}$ 815~keV transition in $^{62}$Fe, populated only in the decay of the low-spin state in $^{62}$Mn~\cite{Pauwels2014,NNDC}, was not observed in coincidence with the $2^{+}_{1}\to0^{+}_{1}$ 877~keV transition or in the singles spectra.
Upper limits of $<3.2\%$ and $<1.8\%$ (2$\sigma$) with respect to the $4^{+}_{1}\to2^{+}_{1}$ 1299~keV transition are determined in the coincidence and singles spectra, respectively.
This supports our analysis that the content of $^{62}$Mn($1^{+}$) in the beam was negligible with respect to the higher-spin state.

\begin{figure}[tb]
\centering
\includegraphics[width=\columnwidth]{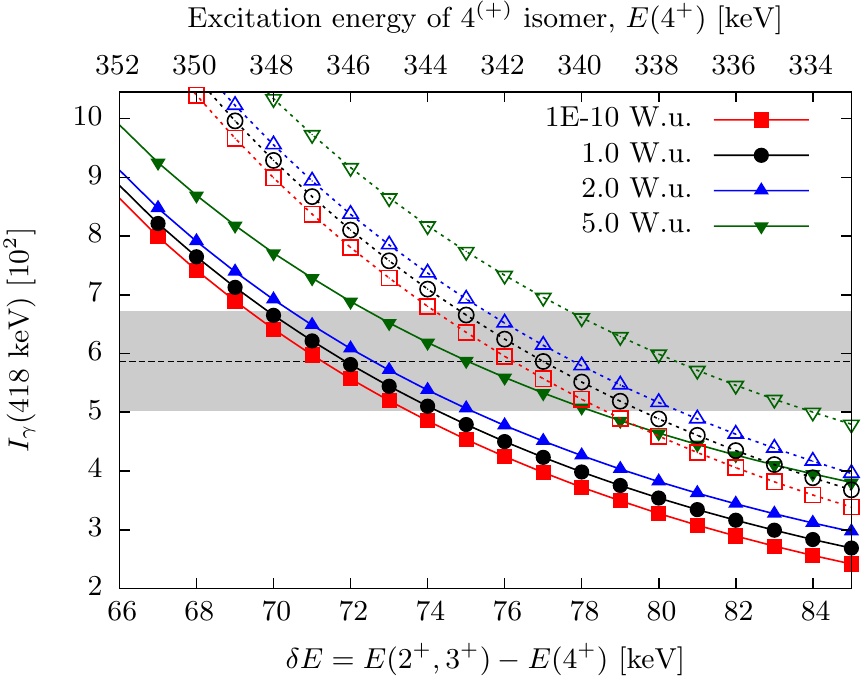}
\caption{
Intensity of the 418~keV $\gamma$-ray transition, calculated in \gosia{}, as a function of energy difference between the $4^{+}$ and the newly proposed states, for different values of $B(E2)\uparrow$.
Filled symbols and solid lines show calculations assumed a $2^{+}$ state, while open symbols and dashed lines are for a $3^{+}$ state.
It is assumed that the conversion coefficient for the transition is negligible and the branching ratio to the $1^{+}$ ground state is 100\%.
The experimentally determined intensity and associated uncertainty (including a contribution from the normalisation to the $^{109}$Ag target excitation) are shown by the horizontal dashed line and shaded area, respectively}
\label{fig:yield_vs_energy}
\end{figure}

\begin{figure}[tb]
\centering
\includegraphics[width=0.96\columnwidth]{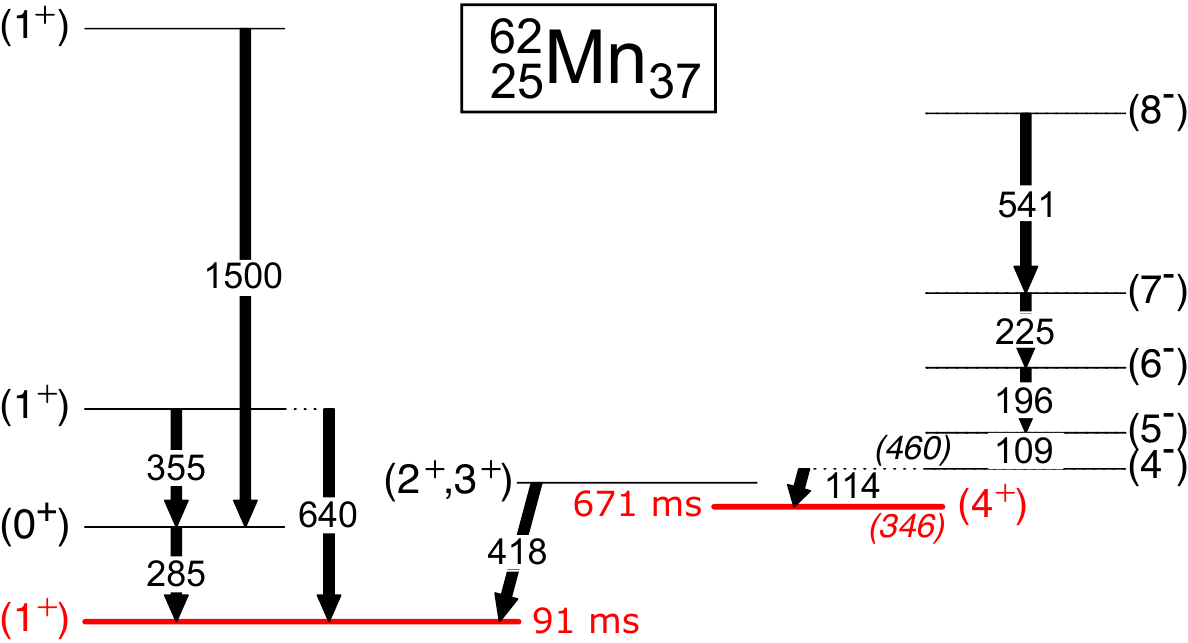}
\caption{
Proposed level scheme of $^{62}$Mn, following the placement of the new $(2^{+},3^{+})$ level at 418(2)~keV.
The position of this state is taken to be $72_{-3}^{+8}$~keV above the $\beta$-decaying $4^{+}$ state.
This reflects the assumption that $I^{\pi}=2^{+}$ with the possibility that $I^{\pi}=3^{+}$.
In the latter case there would be a 5~keV increase in this energy difference, effectively lowering the energy of the $\beta$-decaying $4^{+}$ state to 341~keV.
Both $\beta$-decaying states are shown in red along with their measured halflives~\cite{Hannawald1999,Sorlin1999,Gaudefroy2005}.
Low-spin states decaying to the $\beta$-decaying $1^{+}$ state were determined in Ref.~\cite{Gaudefroy2005}.
High-spin states decaying to the $\beta$-decaying $4^{+}$ state were determined in Refs.~\cite{Valiente-Dobon2008,Chiara2010}.
The arrows represent all $\gamma$ rays that have been observed in previous experiments, including this work, with their energies labelled in keV.
}
\label{fig:62Mn-levels}
\end{figure}

The simplest assumption of the origin of the 418~keV $\gamma$-ray is of de-excitation of a state that is excited via a single 418~keV $E2$ transition from the $\beta$-decaying $4^{+}$ state.
By normalising the intensity of this peak using the same method as in Equation~\ref{eq:beamimpurities}, but now with $F_{\mathrm{Mn}}=1-F_{\mathrm{Fe}}$, it is possible to extract the Coulomb-excitation cross section and hence a $B(E2)$ value.
There are range of spin possibilities for such a state ($I=2,3,4,5,6$).
However, it can be shown that the $B(E2){\uparrow}$ value is relatively independent of the assumption of the spin of the final state and in all cases results in a value of $\approx30$~W.u., more than double that observed for the ${2^{+}_{1} \to 0^{+}_{1}}$ transition in $^{62}$Fe.
This large $B(E2)$ value would imply an intrinsic quadrupole moment, $Q_{2}(4^{+})\approx122$~$e$fm$^{2}$, within the rigid-rotor model~\cite{BohrMottelson}, which can be compared to the limits extracted from the spectroscopic quadrupole moment measured in recent laser-spectroscopy measurements, \linebreak $|Q_{2}(4^{+})|<40$~$e$fm$^{2}$~\cite{Heylen2015,Heylen2015a}.
Such a strong transition is therefore not expected. In order to reproduce the observed Coulomb-excitation cross section in the limit of the measured quadrupole moment, the excitation energy of the state must be lower.
A scenario similar to ``forced isomer depopulation''~\cite{Stefanescu2007} is therefore envisaged, such that the final state of the $\gamma$-ray decay is the $1^{+}$ ground-state proceeding via the excitation of an intermediate $I^{\pi}=(2^{+},3^{+})$ state.
In order to estimate the position of this new state relative to the $\beta$-decaying $4^{+}$ state, it is assumed that they are coupled by a weak $E2$ transition of 1~W.u. in strength.
The sensitivity to this assumption is tested by also taking a negligibly small value of $10^{-10}$~W.u. and a large value of 5~W.u..
It can be shown that the $\gamma$-ray intensity follows roughly linearly as a function of $B(E2)$ for small values ($<10$~W.u.).
The energy of the state is varied in each case and the \gosia{} code is used to calculate the expected $\gamma$-ray intensity.
For this purpose, it is assumed that there is a 100\% branch to the $1^{+}$, following the systematics of $2^{+}$ states in $^{58,60}$Mn~\cite{Steppenbeck2010} and the energy factor in $\gamma$-ray decay, and that the electron conversion can be considered negligible.
While the $3^{+}$ states in $^{58,60}$Mn have a non-negligible branch to the $2^{+}$, it is expected that we populate the state that is lowest in energy and therefore a 100\% branch is also assumed for the $3^{+}$ possibility.
The results of such calculations are plotted in Figure~\ref{fig:yield_vs_energy}, where it is shown that the most likely position of the new state is $\approx72(3)$~keV above the $\beta$-decaying $4^{+}$ state, should it have ${I^{\pi}=2^{+}}$, or $\approx77(3)$~keV for ${I^{\pi}=3^{+}}$.
Here, the assumption of a single-particle transition is taken, such that ${B\left(E2;4^{+}\to\left(2^{+},3^{+}\right)\right)=1(1)}$~W.u..
An updated level \linebreak scheme showing the placement of the new $(2^{+},3^{+})$ state is shown in Figure~\ref{fig:62Mn-levels}.
In turn, the $\beta$-decaying $4^{+}$ is now proposed to lie $346^{+3}_{-8}$~keV above the $\beta$-decaying $1^{+}$, thus solving the conundrum of which $\beta$-decaying state in $^{62}$Mn is the ground state.

\begin{figure}[tb]
\centering
\includegraphics[width=0.98\columnwidth]{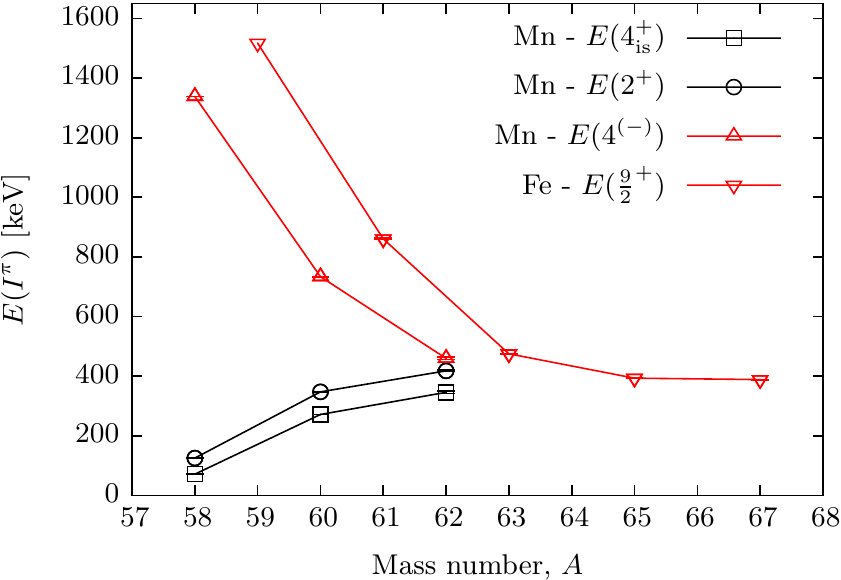}
\caption{Energy systematics of selected levels in the Mn isotopes around $N=36$, with respect to the ground state.
The data for each isotope is taken from the following references:
$^{58}$Mn~\cite{Nesaraja2010},
$^{59}$Fe~\cite{Warburton1977},
$^{60}$Mn~\cite{Steppenbeck2010,Browne2013},
$^{61}$Fe~\cite{Hoteling2010,Radulov2014},
$^{62}$Mn~\cite[and the present work]{Valiente-Dobon2008,Chiara2010},
$^{63}$Fe~\cite{Radulov2014},
$^{65}$Fe~\cite{Radulov2014}, 
$^{67}$Fe~\cite{Radulov2014}.
The assumption of $I=2^{+}$ is taken here for the new state at 418~keV in $^{62}$Mn.
}
\label{fig:Mn-sys}
\end{figure}

To interpret these newly-positioned states, their energies are compared to similar states in the neighbouring isotopes and isotones in Figure~\ref{fig:Mn-sys}.
In both $^{58}$Mn and $^{60}$Mn there are a number of positive-parity states that lie above the isomeric $4^{+}$ state, the first of which are $I=2^{+}$ states at 54~keV and 76~keV, respectively~\cite{Steppenbeck2010}.
This represents a good candidate for the state that we observe in $^{62}$Mn from both its energy and decay behaviour. An inversion of the the $I=2^{+}$ and $3^{+}$ states cannot be ruled out, however.
Low-lying, low-spin positive-parity states were also observed by Gaudefroy~\textit{et~al.}~\cite{Gaudefroy2005}, fed in the $\beta$ decay of $^{62}$Cr.
With the current placement of the $\beta$-decaying $4^{+}$ state at an energy of 346~keV, we are able to conclude that the state at 285~keV must have $I=0$. The alternative scenario of Ref.~\cite{Gaudefroy2005} that sees this state have $I=2$ would not maintain the isomeric nature of the $4^{+}$ due to the potential 59~keV depopulating $E2$ transition.
The importance of the $\nu g_{9/2}$ orbital for $N \ge 34$ Mn isotopes was investigated in shell-model calculations in the $fpg_{9/2}$ valence space~\cite{Srivastava2010}.
The spin of the ground state is incorrectly predicted to be $I=2^{+}$ in all isotopes studied, and the lowest-lying $0^{+}$ state in $^{62}$Mn is predicted to be at 897~keV, more than 600~keV higher than the newly assigned $0^{+}$ at 285~keV.
The lack of an isomeric $4^{+}$ state in these calculations could point towards the importance of $\pi pf$ shell.
Calculations using the new LNPS interaction~\cite{Lenzi2010} have shown how excitations across the $Z=28$ shell closure are key to the evolution of the neutron shells in this region.
Important new experimental developments with COLLAPS at ISOLDE are driving theory in this direction~\cite{Heylen2015,Heylen2015a}.

Chiara~\textit{et~al.}~\cite{Chiara2010} concluded that the state that lies 114~keV above the $\beta$-decaying $4^{+}$ state is the $I=4$ member of the $\pi f_{7/2}^{-1} \nu g_{9/2}^{+1}$ multiplet.
The members of this multiplet would range from $1 \le I \le 8$ and have negative parity, with the lowest- and highest-spin members lying higher in energy than the intermediate-spin states, hence the reason the lowest spins are not observed in the yrast-biased reactions of Refs.~\cite{Valiente-Dobon2008,Chiara2010}.
The proposed $g_{9/2}$ neutron configuration in the Fe isotones gives rise to the yrast $\frac{9}{2}^{+}$ state that is observed to decrease in energy with respect to the ground-state configuration with increasing neutron number.
This behaviour is shown in Figure~\ref{fig:Mn-sys} alongside the corresponding energies of $4^{(-)}$ states from the proposed $\pi f_{7/2}^{-1} \nu g_{9/2}^{+1}$ multiplet in the Mn isotopes.
In the shell-model calculations of Ref.~\cite{Srivastava2010}, the presence of negative-parity states beginning at 424~keV in $^{62}$Mn is consistent with the current placing of the $4^{(-)}$ state.

\section{Summary and conclusions}

In summary, Coulomb excitation has been performed on a mixed beam of $^{62}$Mn/Fe at REX-ISOLDE.
The iron component, which cannot be extracted from the primary ISOL target in conventional means, was produced in-beam via the $\beta$ decay of the manganese parent.
The ratio of iron and manganese was controlled by adjusting the trapping and charge-breeding times of the REX-TRAP/EBIS coupling before the post-accelerator and measured in a gas-filled ionisation chamber.
From this, we were able to provide an independent measurement of the $B(E2; 2^{+}_{1} \to 0^{+}_{1})$ value in $^{62}$Fe under the assumption that $Q_{s}(2^{+}_{1})=0$, previously measured via lifetimes obtained with the recoil-distance Doppler-shift method.
Combining the complementary lifetime and Coulomb-excitation data, we were able to perform a first measurement of the spectroscopic quadrupole moment of the first-excited $2^{+}$ state, albeit with a large uncertainty.
Additionally, a new transition at 418(2)~keV has been observed in $^{62}$Mn and is proposed to arise from the decay of a $I=(2^{+},3^{+})$ state, lying just $72_{-3}^{+8}$~keV above the $\beta$-decaying $4^{+}$ state, to what would be the $1^{+}$ ground state.
This is the first experimental indication of the relative positions of the $1^{+}$ and $4^{+}$ states in $^{62}$Mn, crucial for understanding the role of the $\nu g_{9/2}$ orbital in shell-model calculations.

\begin{acknowledgement}
We acknowledge the support of the ISOLDE Collaboration and technical teams.
This work was supported 
by FWO Vlaanderen GOA/2010/10 (BOF KULeuven), 
by the IAP Belgian Science Policy (BriX network P6/23 and P7/12), 
by BMBF under contract 05P12PKFNE, 
L.P.G. acknowledges support from FWO-Vlaanderen (Belgium) via an FWO Pegasus Marie Curie Fellowship.
\end{acknowledgement}

\bibliography{MnFeCoulex_20151027.bbl}

\end{document}